arXiv: 0909.3616

# Heat capacity of the generalized two-atom and many-atom gas in nonextensive statistics

Guo Lina, Du Jiulin

*Department of Physics, School of Science, Tianjin University, Tianjin 300072, China*

**Abstract**

We have used the generalized two-atom ideal gas model in Tsallis statistics for the statistical description of a real gas. By comparing the heat capacity with the experimental results for the two-atom molecule gases such as $N_2$, $O_2$ and CO, we find that these gases appear extensive at normal temperature, but they may be nonextensive at the lower temperature. Furthermore, we study the heat capacity of the generalized many-atom gas model. We conclude that, for the many-atom gas with a high degree of freedom, a weak nonextensivity of $1-q<0$ can lead to the instability.

## 1. Introduction

In Boltzmann-Gibbs (BG) statistics, the equipartition of energy means that each degree of freedom contributes $\frac{1}{2}k_BT$ to the internal energy where $k_B$ is Boltzmann constant. The degree of freedom of the one-atom ideal gas is three, while that of the two-atom one is five, including three degrees of freedom for the motion of the center of mass and two degrees of freedom for the rotation of two atoms around their center of mass. So the internal energy of two-atom ideal gas with N molecules is $\frac{5}{2}Nk_BT$. According to the relation between the internal energy and the heat capacity

$$C_v = \frac{dE}{dT} \qquad (1)$$

we have the heat capacity of the two-atom ideal gas,

$$C_v = \frac{d}{dT}\frac{5}{2}Nk_BT = \frac{5}{2}Nk_B \qquad (2)$$

This result is obtained based on the assumption of ideal gas, in which there is not any interaction between the molecules, and therefore it can be used only for the two-atom gas where the interactions are so small that they can be neglected. But if we want to use Eq.(2) for the gases with long-range interactions, such as the plasma gas and the self-gravitating gas where the inter-particle interactions can not be neglected, we should make the modification in Eq.(2) in order that the effects of the interactions might be taken into account. Nonextensive statistics may be applied to study the effects.

Tsallis proposed the so-called q-entropy in 1988 as the nonextensive generalization of BG entropy. The q-entropy formula [1] is

$$S_q = -k_B \frac{1 - \sum_i p_i^q}{(q-1)} \qquad (3)$$

where $p_i$ is the probability of the $i$th microstate and $q$ is a parameter different from unity, quantifying the degree of nonextensivity. In the limit $q \to 1$, the entropy recovers to the BG extensive formula, $S = -k_B \sum_i p_i \ln p_i$. The statistics based on Tsallis' q-entropy has been called nonextensive statistics. It has been shown that the kinetic foundations of nonextensive statistics lead to the power-law q-distribution function [2, 3],

$$f = B_q \left[1 - (1-q)\beta E\right]^{\frac{1}{1-q}} \qquad (4)$$

where $\beta = \frac{1}{k_B T}$, $E$ is the energy of the particle. For one-atom particle, the energy contains only kinetic energy, representing the motion of the centre of mass, which is

$$E = \frac{p^2}{2m} = \frac{1}{2}mv^2 \qquad (5)$$

Therefore, the q-distribution function can be written by

$$f(p) = B_q \left[1 - (1-q)\beta \frac{p^2}{2m}\right]^{\frac{1}{1-q}} \qquad (6)$$

Nonextensive statistical theory has been applied to so many interesting fields, such as astrophysics (see [4] and the references therein, [5] ), real gases [6][17], plasma [7], nuclear reactions [8] and so on. It was also applied to study the nonextensive effects introduced by the potential with those including long-range interactions in the nonequilibrium plasma and self-gravitating systems. In the description, the effects of the long-range interactions are replaced by introducing the nonextensive parameter $q \neq 1$ [9-13]. For example, we have known now that the nonextensive parameter $q$ can be expressed by the inter-particle interacting potential $\varphi$ as the formulae $k_B \nabla T + (1-q) m \nabla \varphi = 0$ for the self-gravitating gases [10, 11] and $k_B \nabla T - (1-q) e \nabla \varphi = 0$ for the plasma gases [12], which finds the experimental support from the helioseismological measurements [13].

For the classical ideal gas model, it is assumed that there is not any interaction between the particles. However, in fact, there must be the gravitation between them, although it is too week to be considered. While in the low temperature, the kinetic energy of the particle is much smaller than in the normal temperature, so the potential energy, which is mainly offered by the gravitation, becomes more important. At this time, the property of a gas has been some departure from the ideal gas model. In this paper, we try to use the generalized gas model in Tsallis statistics to describe the many-atom gas, taking into account those including the gravitational interactions between particles in the gas.

In Sec.2, we study the heat capacity of a two-atom molecule gas using the generalized gas model in Tsallis statistics. In Sec.3, we compare the present results with the experimental data for the heat capacity of the gases $N_2$, $O_2$ and CO, respectively. In Sec.4, we study the contribution of the degree of freedom and the nonextensivity to the heat capacity for a many-atom molecule gas. Finally in Sce.5, we give our conclusions.

## 2．Heat capacity of the generalized gas with rigid two-atom molecules

First, we consider the generalized ideal gas composed of rigid two-atom molecules to study the heat capacity. This means that we don't care about the vibration motion between the two atoms. Due to two-atom molecule cannot be considered as one mass point, its behavior contains two parts: one is the rotation around the center of mass and the other is the motion of the center of mass in 3D space. Therefore, we must take into account the contribution of the rotation to internal energy when we study the heat capacity using the new statistics.

Two atoms in the same two-atom molecule can be considered as two mass points. Their mass is $m_1, m_2$ respectively. The distance between the two atoms is $l$. Then the rotational kinetic energy of the rotation around the center of mass [14] is written as

$$E_1 = \frac{1}{2\phi}\left(p_\theta^2 + \frac{1}{\sin^2\theta}p_\varphi^2\right) \tag{7}$$

where $\phi = \mu l^2$ is the rotational inertia and $\mu = \dfrac{m_1 m_2}{m_1 + m_2}$ is the reduced mass of two atoms. $\theta$, $\varphi$ is zenith and azimuth respectively. $p_\theta$ and $p_\varphi$ is the generalized momentum in the degree of freedom of $\theta$ and $\varphi$. The kinetic energy of the center of mass is written as

$$E_2 = \frac{1}{2m}\left(p_x^2 + p_y^2 + p_z^2\right) \tag{8}$$

where $m = m_1 + m_2$. Therefore, the total energy of a two-atom molecule is

$$E = \frac{1}{2m}\left(p_x^2 + p_y^2 + p_z^2\right) + \frac{1}{2\phi}\left(p_\theta^2 + \frac{1}{\sin^2\theta}p_\varphi^2\right) \tag{9}$$

Substituting Eq. (9) into (4), we can get the q-distribution function of the generalized two-atom molecule gas,

$$f(p) = B_q\left\{1-(1-q)\beta\left[\frac{1}{2m}\left(p_x^2 + p_y^2 + p_z^2\right) + \frac{1}{2\phi}\left(p_\theta^2 + \frac{1}{\sin^2\theta}p_\varphi^2\right)\right]\right\}^{\frac{1}{1-q}} \tag{10}$$

The q-expectation value of energy in Tsallis statistics is defined as

$$\langle E \rangle_q = \frac{\int_{-p_{xm}}^{p_{xm}}\int_{-p_{ym}}^{p_{ym}}\int_{-p_{zm}}^{p_{zm}}\int_{-p_{\theta m}}^{p_{\theta m}}\int_{-p_{\varphi m}}^{p_{\varphi m}} f^q E \, dp_\varphi dp_\theta dp_z dp_y dp_x}{\int_{-p_{xm}}^{p_{xm}}\int_{-p_{ym}}^{p_{ym}}\int_{-p_{zm}}^{p_{zm}}\int_{-p_{\theta m}}^{p_{\theta m}}\int_{-p_{\varphi m}}^{p_{\varphi m}} f^q \, dp_\varphi dp_\theta dp_z dp_y dp_x}$$

$$= \frac{\int_{-p_{xm}}^{p_{xm}}\int_{-p_{ym}}^{p_{ym}}\int_{-p_{zm}}^{p_{zm}}\int_{-p_{\theta m}}^{p_{\theta m}}\int_{-p_{\varphi m}}^{p_{\varphi m}}\left\{1-(1-q)\beta\left[\frac{1}{2m}\left(p_x^2+p_y^2+p_z^2\right)+\frac{1}{2\phi}\left(p_\theta^2+\frac{1}{\sin^2\theta}p_\varphi^2\right)\right]\right\}^{\frac{q}{1-q}}\left[\frac{1}{2m}\left(p_x^2+p_y^2+p_z^2\right)+\frac{1}{2\phi}\left(p_\theta^2+\frac{1}{\sin^2\theta}p_\varphi^2\right)\right]dp_\varphi dp_\theta dp_z dp_y dp_x}{\int_{-p_{xm}}^{p_{xm}}\int_{-p_{ym}}^{p_{ym}}\int_{-p_{zm}}^{p_{zm}}\int_{-p_{\theta m}}^{p_{\theta m}}\int_{-p_{\varphi m}}^{p_{\varphi m}}\left\{1-(1-q)\beta\left[\frac{1}{2m}\left(p_x^2+p_y^2+p_z^2\right)+\frac{1}{2\phi}\left(p_\theta^2+\frac{1}{\sin^2\theta}p_\varphi^2\right)\right]\right\}^{\frac{q}{1-q}}dp_\varphi dp_\theta dp_z dp_y dp_x}$$

$$\tag{11}$$

where $p_{xm}, p_{ym}, p_{zm} = \sqrt{\dfrac{2m}{(1-q)\beta}}$, $p_{\theta m} = \sqrt{\dfrac{2\phi}{(1-q)\beta}}$, $p_{\varphi m} = \sqrt{\dfrac{2\phi\sin^2\theta}{(1-q)\beta}}$ is a thermal cutoff on the maximum value allowed for the momentum of particles if $q < 1$, whereas without cutoff if $q>1$, $p_m \to \infty$. If $q<1$, integrate Eq. (11) we can get

$$\langle E \rangle_q = \frac{5}{(7-5q)\beta} = \frac{5}{7-5q}k_B T \tag{12}$$

Then using Eq. (1) we obtain the heat capacity,

$$C_V = \frac{5}{7-5q}Nk_B, \quad q < \frac{7}{5}, \tag{13}$$

where $N$ is the number of particles. When $q \to 1$, Eq.(13) becomes the standard form $C_V = \frac{5}{2}Nk_B$. When $q < \frac{7}{5}$, $C_v > 0$. When $q \to 7/5$, the heat capacity diverges ($C_v \to \infty$).

3. **Comparing the result with experiments**

Table I gives the experimental values of $\gamma = \frac{C_p}{C_v}$ and the heat capacities at constant pressure for the gases $N_2$, $O_2$ and CO, respectively [14]. The heat capacities at constant volume can be obtained by $\gamma = C_p/C_v$, and then we use Eq(13) to find out the values of parameter $q$.

| Gas | $T(K)$ | $\gamma = C_p/C_v$ | $C_p/Nk_B$ | $C_v/Nk_B$ | $q$ |
|---|---|---|---|---|---|
| $N_2$ | 293 | 1.398 | 3.51 | 2.51 | 1.00 |
|  | 92 | 1.419 | 3.38 | 2.38 | 0.98 |
| $O_2$ | 293 | 1.398 | 3.51 | 2.51 | 1.00 |
|  | 197 | 1.411 | 3.43 | 2.43 | 0.99 |
|  | 92 | 1.404 | 3.47 | 2.47 | 0.99 |
| CO | 291 | 1.396 | 3.52 | 2.52 | 1.00 |
|  | 93 | 1.417 | 3.40 | 2.40 | 0.98 |

Table I. The experimental values of $C_v$, $C_p$, $\gamma$ and the corresponding q.

From the table we find that the values of the parameter $q$ for three gases are equal to 1 at normal temperature but at lower temperature are deflected away from 1. This deviation could not be considered as the error in the experimental measurements. Otherwise, some of them should be bigger than unity and some of them should be smaller than unity and their average value should be unity. So, the deviation from the unity of the data in table 1 should be considered as the nonextensive effect. These experimental values coming from the textbook clearly show the property that the gases appear *extensive* at normal temperature but they may be *nonextensive* at the low temperature. We understand that kinetic energy of a gas is much smaller at low temperature than at normal temperature, as compared with potential energy, so the nonextensive effect produced by the long-range interactions of inter-particle, such as gravitation, becomes important at low temperature. In other words, the generalized ideal gas model in $q$-statistics is more close to the real gas. Especially at low temperature, we should use the generalized two-atom ideal gas model for the statistical description of gases. Thus, when the long-range interactions between particles becomes important, we can use the generalized ideal gas model in Tsallis statistics as a substitute of ideal gas model in BG statistics to study the statistical property of a real gas.

Besides at lower temperature, we suppose that the heat capacity of the gases at higher pressure should also be nonextensive because the interaction between particles in the gases should become more important at higher pressure, as if those at lower temperature, but we still need the data of the experimental evidences.

### 4. The contribution of degree of freedom to the heat capacity

In BG statistics, the theorem of equipartition of energy tells us that each degree of freedom contributes $\frac{1}{2}k_B T$ to internal energy. Some authors have analyzed this theorem in new framework of Tsallis statistics [15]. We consider a one-atom molecule gas with 3 degrees of freedom, responding to the motion in $x, y, z$ direction. We have the q-expectation value of energy,

$$\langle E \rangle_q = \left\langle \frac{1}{2}mv^2 \right\rangle_q = \frac{\frac{m}{2}\int_{-v_m}^{v_m} f^q v^2 d^3v}{\int_{-v_m}^{v_m} f^q d^3v}, \tag{14}$$

with the distribution function $f(v) = B_q \left[1 - (1-q)\beta \frac{mv^2}{2}\right]^{\frac{1}{1-q}}$. Integrate Eq.(14) (see the appendix), we find

$$\frac{3}{2} - \beta \langle E \rangle_q = \frac{3}{2}(1-q)\beta \langle E \rangle_q \tag{15}$$

and then,

$$\langle E \rangle_q = \left\langle \frac{1}{2}mv^2 \right\rangle_q = \frac{3}{(5-3q)\beta} \tag{16}$$

We consider a two-atom molecule gas with 5 degrees of freedom, including the motion in $x, y, z$ direction and the rotation around centre of mass. The q-expectation value of energy is written by

$$\langle E \rangle_q = \frac{\int_{-p_{xm}}^{p_{xm}} \int_{-p_{ym}}^{p_{ym}} \int_{-p_{zm}}^{p_{zm}} \int_{-p_{\theta m}}^{p_{\theta m}} \int_{-p_{\varphi m}}^{p_{\varphi m}} f^q E dp_\varphi dp_\theta dp_z dp_y dp_x}{\int_{-p_{xm}}^{p_{xm}} \int_{-p_{ym}}^{p_{ym}} \int_{-p_{zm}}^{p_{zm}} \int_{-p_{\theta m}}^{p_{\theta m}} \int_{-p_{\varphi m}}^{p_{\varphi m}} f^q dp_\varphi dp_\theta dp_z dp_y dp_x} \tag{17}$$

where $E$ is the kinetic energy of each particle, given by Eq.(9). Integrate Eq.(17), we find

$$\frac{5}{2} - \beta \langle E \rangle_q = \frac{5}{2}(1-q)\beta \langle E \rangle_q \tag{18}$$

and then

$$\langle E \rangle_q = \frac{5}{(7-5q)\beta} \tag{19}$$

We now consider a N-atom molecule gas with $g$ degrees of freedom and its kinetic energy can be written as $E = a_1 p_1^2 + a_2 p_2^2 + \cdots + a_g p_g^2$, where $p_1, p_2 \cdots p_g$ is called the generalized momentum. Then the q-expectation value of energy is

$$\langle E \rangle_q = \frac{\int_{-p_{1m}}^{p_{1m}} \int_{-p_{2m}}^{p_{2m}} \cdots \int_{-p_{gm}}^{p_{gm}} f^q E dp_g \cdots dp_2 dp_1}{\int_{-p_{1m}}^{p_{1m}} \int_{-p_{2m}}^{p_{2m}} \cdots \int_{-p_{gm}}^{p_{gm}} f^q dp_g \cdots dp_2 dp_1}. \tag{20}$$

Integrating Eq.(20) we find (see the appendix)

$$\frac{g}{2} - \beta \langle E \rangle_q = \frac{g}{2}(1-q)\beta \langle E \rangle_q, \tag{21}$$

and then,

$$\langle E \rangle_q = \frac{g}{2 + g(1-q)} k_B T \tag{22}$$

Therefore, the heat capacity is

$$C_V = \frac{g}{2 + g(1-q)} N k_B \tag{23}$$

So, the heat capacity of the generalized ideal gas model is generally related with the degree of freedom and the nonextensive parameter $q$. If $q \to 1$, $C_V$ recovers to the standard form in BG statistics, $\frac{g}{2}Nk_B$. Obviously, if $q=1$, which correspond with BG statistics, $C_V/Nk_B$ increases with the change of $g$ linearly. If $q<\frac{2}{g}+1$, $C_V/Nk_B$ is always positive for any $g$ and increase with $g$ continuously. While if $q\geq\frac{2}{g}+1$, $C_V/Nk_B$ will diverge, leading to a new relation between the degree of freedom $g$ and the nonextensive parameter $q$ that $q$ must be less than $\frac{2}{g}+1$ so that the capacity is finite.

For example, if the generalized gas is with rigid many-atom molecule, the motion may contain two parts: one is the motion of the centre of mass, the other is the relative motion between atoms. For the rigid molecule, the distance between atoms is constant, so the molecule can be considered as a rigid body and the relative motion between atoms becomes a rotation of the rigid body. Therefore, the energy of each molecule is written [14] as

$$E = \frac{1}{2m}\left(p_x^2 + p_y^2 + p_z^2\right) + \sum_{i=1}^{3}\frac{L_i^2}{2\phi_i} \tag{24}$$

where $\phi_i$ is the rotational inertia, describing the rotation around $i$ th principal axis of inertia, and $L_i$ is the projection of angular momentum on $i$ th principal axis. The degrees of freedom of such a many-atom molecule is $g=6$. Using Eq.(23), we get the heat capacity,

$$C_V = \frac{6}{8-6q}Nk_B \tag{25}$$

If $q\to 1$, $C_v$ is $3Nk_B$, recovered to the form in BG statistics. But if $q\neq 1$, $C_v$ will be different from $3Nk_B$, the nonextensive effect introduced by the gravitational interactions becomes important.

Fig.1 shows the heat capacity $C_v$ changes with the parameter $1-q$ when the degree of freedom takes $g=1,3,5$, respectively. We find that $C_v$ grows continuously with decrease of $(1-q)$ for any $g$. In particular, when $1-q<0$, $C_v$ will increase very rapidly with decrease of $(1-q)$ and diverge to become infinite. The more the degree of freedom $g$ is, the more speed $C_v$ diverges, if only 1-$q$ < 0. The infinite heat capacity of a gas is known to correspond to its instability [16]. This means that the heat capacity of a many-atom gas with a *high* degree of freedom would become infinite very easily if the gas is nonextensive and $1-q<0$, *although its nonextensivity may be very weak*, and consequently the gas would be unstable. We therefore conclude that, for a many-atom gas with a *high* degree of freedom, the weak nonextensivity of $1-q<0$ can lead to its instability.

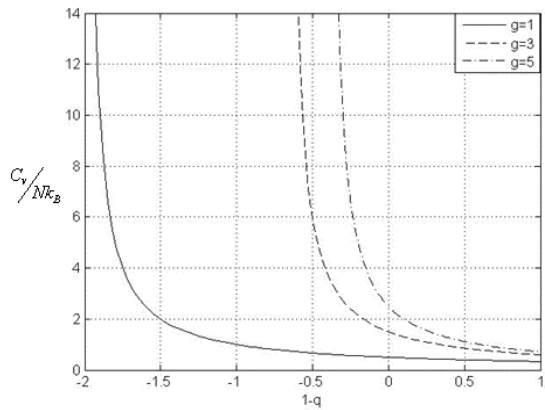

Fig1. The change of $C_v$ for different degree of freedom $g$

Fig. 2 illustrates the heat capacity changes with the degree of freedom $g$ for several different values of q=0.5, 0.8, 1, 1.2, 1.5, respectively, from the other point of view, showing the same property as above of the gas.

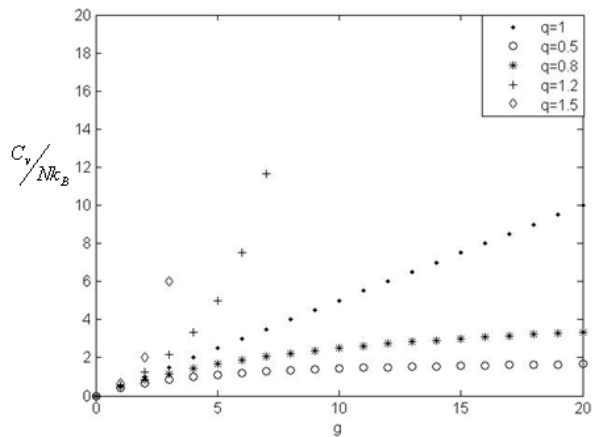

Fig 2. The change of $C_v$ for different q

## 5. Conclusion

In this paper, we have used the generalized ideal gas model in Tsallis statistics for the statistical description of a real gas with the long-range inter-particle interactions. First of all, we consider the generalized two-atom molecule gas to get the q-expectation value of energy, and then we obtain the heat capacity Eq.(13) that depends on the degree of freedom and the nonextensive parameter $q$. Second, we compare the conclusion with the experimental results for the two-atom molecule gases such as $N_2$, $O_2$ and CO, respectively. We find that these gases appear *extensive* at normal temperature, but they may be *nonextensive* at the low temperature. So we understand that the generalized ideal gas model in Tsallis q-statistics may be more suitable for the statistical description of a real gas when consider the weak long-range inter-particle interactions. We can suppose that the heat capacity of the gases at higher pressure be also nonextensive because in such a physical situation the weak inter-particle interactions become more important than at normal pressure.

Furthermore, we have studied the heat capacity of the generalized gas with the degree of freedom, denoted by $g$. We derive a formula Eq. (23) where the heat capacity is generally related with the degree of freedom $g$ and the nonextensive parameter $q$. If taking $q \to 1$, Eq.(23) just becomes the theorem of equipartition of energy in BG statistics. We also find that the degree of freedom $g$ and the nonextensive parameter $q$ must satisfy $q < \frac{2}{g} + 1$ if the capacity is finite.

In Fig 1 and Fig 2, we show the numerical relation between the heat capacity, the nonextensive parameter $q$ and the degree of freedom $g$. We conclude that, for a many-atom gas with a *high* degree of freedom, the *weak nonextensivity* of $1 - q < 0$ can lead to its instability. As an example, we calculate the heat capacity of the many-atom molecule gas with six degrees of freedom.

*Additional remarks*: Recently, [17] study the relation of temperature of the rigid gas and particle number to the heat capacity. [18][19] deal with the related topics to Tsallis statistics.

**Appendix**

The energy can be written by

$$E = a_1 p_1^2 + a_2 p_2^2 + \cdots + a_g p_g^2 \tag{A1}$$

where $p_1, p_2 \cdots p_g$ is generalized momentum. Then the q-distribution function is

$$f(p) = B_q \left[ 1 - (1-q)\beta \left( a_1 p_1^2 + a_2 p_2^2 + \cdots + a_g p_g^2 \right) \right]^{\frac{1}{1-q}} \tag{A2}$$

The q-expectation value of energy is

$$\langle E \rangle_q = \frac{\int_{-p_{1m}}^{p_{1m}} \int_{-p_{2m}}^{p_{2m}} \cdots \int_{-p_{gm}}^{p_{gm}} f^q E dp_g \cdots dp_2 dp_1}{\int_{-p_{1m}}^{p_{1m}} \int_{-p_{2m}}^{p_{2m}} \cdots \int_{-p_{gm}}^{p_{gm}} f^q dp_g \cdots dp_2 dp_1} \tag{A3}$$

where $p_{im} = \sqrt{\frac{1}{(1-q)\beta a_i}}$ (i=1, 2,$\cdots$g) is a thermal cutoff on the maximum value allowed for the momentum of the particles ($q<1$), whereas for the power law q-distribution without cutoff ($q>1$) $p_m \to \infty$.

Substituting Eq.(A1) and (A2) into (A3), we have

$$\langle E \rangle_q = \frac{\int_{-p_{1m}}^{p_{1m}} \int_{-p_{2m}}^{p_{2m}} \cdots \int_{-p_{gm}}^{p_{gm}} \left[ 1 - (1-q)\beta \left( a_1 p_1^2 + a_2 p_2^2 + \cdots + a_g p_g^2 \right) \right]^{\frac{q}{1-q}} \left( a_1 p_1^2 + a_2 p_2^2 + \cdots + a_g p_g^2 \right) dp_g \cdots dp_2 dp_1}{\int_{-p_{1m}}^{p_{1m}} \int_{-p_{2m}}^{p_{2m}} \cdots \int_{-p_{gm}}^{p_{gm}} \left[ 1 - (1-q)\beta \left( a_1 p_1^2 + a_2 p_2^2 + \cdots + a_g p_g^2 \right) \right]^{\frac{q}{1-q}} dp_g \cdots dp_2 dp_1} \tag{A4}$$

Let $u_i = \sqrt{(1-q)\beta a_i} p_i$, $u^2 = u_1^2 + u_2^2 + \cdots + u_g^2$, then it becomes

$$\langle E \rangle_q = \frac{\int_{-1}^{1} (1-u^2)^{\frac{q}{1-q}} \frac{1}{(1-q)\beta} u^2 d^g u}{\int_{-1}^{1} (1-u^2)^{\frac{q}{1-q}} d^g u} \tag{A5}$$

where $u^2$ can be considered as a vector in $g$-D space. All terms in the integral are equivalent, so the space is isotropy. Therefore, Eq.(A5) can be transformed into $g$-D sphere coordinate,

$$\langle E \rangle_q = \frac{\int_0^1 (1-u^2)^{\frac{q}{1-q}} \frac{1}{(1-q)\beta} u^2 A u^{g-1} du}{\int_0^1 (1-u^2)^{\frac{q}{1-q}} A u^{g-1} du} \tag{A6}$$

where $A u^{g-1}$ is the surface area of $g$-D sphere. Let $v = 1 - u^2$, then

$$\langle E\rangle_q = \frac{1}{(1-q)\beta}\frac{\int_0^1 v^{\frac{q}{1-q}}(1-v)^{\frac{g}{2}}dv}{\int_0^1 v^{\frac{q}{1-q}}(1-v)^{\frac{g}{2}-1}dv} \equiv \frac{1}{(1-q)\beta}x \tag{A7}$$

where $\int_0^1 v^{\frac{q}{1-q}}(1-v)^{\frac{g}{2}}dv = (1-q)\frac{g}{2}\int_0^1 v^{\frac{1}{1-q}}(1-v)^{\frac{g}{2}-1}dv$, so

$$x = (1-q)\frac{g}{2}\frac{\int_0^1 v^{\frac{1}{1-q}}(1-v)^{\frac{g}{2}-1}dv}{\int_0^1 v^{\frac{q}{1-q}}(1-v)^{\frac{g}{2}-1}dv} \tag{A8}$$

Therefore

$$\frac{g}{2}(1-q)-x = \frac{g}{2}(1-q)\left[1-\frac{\int_0^1 v^{\frac{1}{1-q}}(1-v)^{\frac{g}{2}-1}dv}{\int_0^1 v^{\frac{q}{1-q}}(1-v)^{\frac{g}{2}-1}dv}\right] = \frac{g}{2}(1-q)x \tag{A9}$$

$$\frac{g}{2}-\beta\langle E\rangle_q = \frac{g}{2}(1-q)\beta\langle E\rangle_q \tag{A10}$$

So we get

$$\langle E\rangle_q = \frac{g}{(2+g-gq)\beta} \tag{A11}$$

## Acknowledgements

This work is supported by the National Natural Science Foundation of China under grant No.10675088.

## References

[1] C. Tsallis, J. Stat. Phys. **52**(1988)479.

[2] R. Silva Jr., J.A.S. Lima, A.R.Plastino, Phys. Lett. A, **249**(1998)401.

[3] R.S. Mendes, C. Tsallis, Phys. Lett. A, **285**(2001)273.

[4] Tsallis C., "Nonextensive Statistical Mechanics-An Approach to Coplexity" (Conference 2007); G. Contopoulos and P. A. Patsis, Chaos in Astronomy, Astrophysics and Space Science Proceedings, P.309-318, Springer Berlin Heidelberg, 2009.

[5] J. M. Silva, R. E. de Souza, J. A. S. Lima, arXiv:0903.0423v1; J. L. Du, New Astron. **12**(2006) 60; J. L. Du, New Astron. **12**(2007)657; J. C. Carvalho, R. Silva, J. D. Jr. do Nascimento, J.

R. DeMedeiros, Europhys. Lett. **84**(2008) 590001.

[6] Y. H. Zheng, J. L. Du, Inter. J. Mod. Phys. B **21**(2007)947;

Liu L Y, Liu Z P and Guo L N, Physica A 387(2008)5768.

[7] J.A.S. Lima, R. Silva Jr. and J. Santos, Phys. Rev. E **61**(2000)3260; J. L. Du, Phys. Lett. A **329**(2004)262; A. Lavagno, P. Quarati, Astrophys. Space Sci. **305**(2006)253; S. Shaikh, A. Khan, P. K. Bhatia, Phys. Lett. A **372**(2008) 1452; L. Y. Liu, J. L. Du, Physica A **387**(2008)4821; A. M. Scarfone, P. Quarati, G. Mezzorani, M. Lissia, Astrophys. Space Sci. **315**(2008)353.

[8] J. L. Wu, X. M. Huang, Chin. Phys. **16**(2007)3216; F.L.M.Pereira, R.Silva, J.S.Alcaniz, Phys.Rev.C **76**(2007)015201; J. L.Wu, H. J. Chen, Phys. Scr. **75**(2007)722; J. L.Wu, H. J. Chen, Mod. Phys. Lett.B **21**(2007)103.

[9] J. L. Du, Astrophys. Space Sci. **312**(2007)47.

[10] J. L. Du, Europhys. Lett. **67**(2004)893.

[11] J. L. Du, Astrophys. Space Sci. **305**(2006)247.

[12] J. L. Du, Phys. Lett. A **329**(2004)262.

[13] J. L. Du, Europhys. Lett. **75**(2006)861.

[14] Q. R. Zhang, *Statistical Mechanics*, Science Press, Beijing, 2004.

[15] A.R. Plastino, J.A.S. Lima, Phys. Lett. A **260**(1999)46.

[16] L.Y. Liu, Z. P. Liu, L. N. Guo, Physica A **387**(2008)5768.

[17] R. Chakrabarti, R. Chandrashekar, S.S. Naina Mohammed, Physica A **387** (2008) 4589.

[18] G. Wilka, Z. Włodarczyk, Physica A **387** (2008) 4809.

[19] N.G. de Almeida, Physica A **387** (2008) 2745.